\documentclass[twocolumn,aps,prl,superscriptaddress,amsfonts]{revtex4-1}

\usepackage{amsmath}
\usepackage{amssymb}
\usepackage{graphicx}
\usepackage{bm}
\usepackage{array}
\usepackage{dcolumn}
\usepackage{hepparticles}
\usepackage{heppennames}
\usepackage{hepnicenames}
\usepackage{epstopdf}

\newcommand{\ket}[1]{{| {#1} \rangle}}

\begin{document}

\title{Radial two-dimensional ion crystals in a linear Paul trap}

\author{Marissa D'Onofrio, Yuanheng Xie, A.J. Rasmusson, Evangeline Wolanski, Jiafeng Cui, and Philip Richerme}
\affiliation{Indiana University Department of Physics, Bloomington, Indiana 47405, USA}
\affiliation{Indiana University Quantum Science and Engineering Center, Bloomington, Indiana 47405, USA}

\date{\today}

\begin{abstract}
We experimentally study two-dimensional (2D) Coulomb crystals in the ``radial-2D" phase of a linear Paul trap. This phase is identified by a 2D ion lattice aligned entirely with the \mbox{radial plane and} is created by imposing a large ratio of axial to radial trapping potentials. Using arrays of \mbox{up to 19} $^{171}$Yb$^+$ ions, we demonstrate that the structural phase boundaries of such crystals are well-described by the pseudopotential approximation, despite the time-dependent ion positions driven by intrinsic micromotion. We further observe that micromotion-induced heating of the radial-2D crystal is confined to the radial plane. Finally, we verify that the transverse motional modes, which are used in most ion-trap quantum simulation schemes, are well-predictable numerically and remain decoupled and cold in this geometry. Our results establish radial-2D ion crystals as a robust experimental platform for realizing a variety of theoretical proposals in quantum simulation and computation.

\end{abstract}

\maketitle
Laser-cooled ions in radio-frequency (rf) and Penning traps form Coulomb crystals, spatially ordered structures that arise due to a balance between trapping fields and Coulomb repulsion. 
Decades of advancements in the preparation and control of cold ion crystals have allowed for the precise manipulation of their internal and external degrees of freedom \cite{drewsen2015ion}, giving rise to applications spanning plasma physics \cite{malmberg1975properties,dubin1999trapped}, high-precision spectroscopy \cite{chou2010frequency, furst2020coherent}, cold molecules \cite{molhave2000formation,rellergert2011measurement,lien2014broadband}, and quantum computation \cite{brown2016co,linke2017experimental,wright2019benchmarking} and simulation \cite{friedenauer2008simulating,richerme2014non,jurcevic2014quasiparticle}. In these experiments, achieving the desired level of control has typically required an initial characterization of ion positions, structural phases, normal mode frequencies, and sources of crystal heating.

Over the last decade, one-dimensional (1D) ion chains in rf traps have seen remarkable success in engineering high-fidelity quantum gates \cite{gaebler2016high,ballance2016high} and simulating 1D quantum spin systems \cite{monroe2021programmable}. 

If a comparable ability to control and probe two-dimensional (2D) crystals in rf traps can be achieved, then the native 2D interactions between ions would provide an inherent advantage over 1D systems for the quantum simulation of complex 2D materials \cite{yoshimura2015creation,richerme2016two,nath2015hexagonal,balents2010spin}. In addition, 2D arrays can hold larger numbers of qubits more efficiently than 1D strings, with
a higher error threshold for fault-tolerance \cite{wang2015quantum,shen2014high}, and may simplify preparations of 2D cluster states for one-way quantum computing \cite{raussendorf2001one,wunderlich2009two}. Already, 2D arrays of ions in Penning traps have led to successes in simulating and studying quantum spin models \cite{britton2012engineered,garttner2017measuring}, though the fast crystal rotation in such traps poses a significant challenge to individual ion addressing.

In rf traps, there are two primary ways to orient a 2D crystal. The first of these, which is an extension of the well-known ``zig-zag" phase, spans a 2D plane defined by one radial and one axial trap direction \cite{sm}. \mbox{In this case,} rf-driven micromotion is present along one of the in-plane directions as well as transverse to the plane. Ion crystals in this phase, which we refer to as the ``lateral-2D" geometry, were first realized in rf traps over 20 years ago \cite{block2000crystalline}. More recent work has measured the vibrational spectrum of lateral-2D crystals \cite{kaufmann2012precise}, and further experiments have demonstrated coherent operations in this regime \cite{wang2020coherently}.

In contrast, the ``radial-2D" phase, defined as the configuration for which the ion plane is coincident with the trap's radial plane, remains largely unexplored experimentally. In this phase, the longitudinal in-plane modes lie along the radial direction and experience micromotion, while the transverse modes lie along the axial direction and remain micromotion-free. This radial-2D phase has been the primary interest for most theoretical studies of 2D ion crystals, which have made predictions of crystal stability, lifetimes, heating rates, phase boundaries, and gate fidelities \cite{yoshimura2015creation,richerme2016two,nath2015hexagonal,buluta2008investigation,buluta2009structure,wang2015quantum,shen2014high}. To date, however, experiments performed with radial-2D crystals have only demonstrated Doppler cooling \cite{ivory2020paul} and probed the radial-2D phase boundary with 3-4 ions \cite{kaufmann2012precise}.

Notably, lateral-2D and radial-2D crystals are each expected to exhibit distinct behavior due to the different relative orientations of micromotion with respect to the crystal plane. Thus, previous studies of the structural and dynamical properties of lateral-2D crystals are not directly applicable to the radial-2D regime \cite{yoshimura2015creation,richerme2016two}.  
 
Moreover, for radial-2D crystals, it is experimentally unknown the degree to which micromotion may obscure site-specific imaging resolution, or worse, lead to fast absorption of energy from the rf drive \cite{ryjkov2005simulations,zhang2007molecular,buluta2008investigation} and melting of the ion lattice \cite{chen2013measurement}.

In this Letter, we report the experimental characterization and coherent control of radial-2D crystals in a linear Paul trap. We map the full range of structural phases for Coulomb crystals as a function of ion number using arrays of up to 19 ions, and we investigate the transverse vibrational mode spectrum in the radial-2D phase. Next, we measure the time-dependent temperature of the crystal as it experiences micromotion-induced heating, and we extract the center-of-mass heating rate along the micromotion-free direction perpendicular to the radial plane. Finally, we discuss the implications for future quantum information processing experiments.

Experiments are performed with $^{171}$Yb$^+$ ions confined in a four-rod linear Paul trap with two ``needle" endcaps along the axial ($\hat{z}$) direction (see supplementary material for detailed trap information \cite{sm}).

A slight asymmetry is introduced between the radial $\hat{x}$- and $\hat{y}$- directions to prevent a zero-frequency rotational mode; for specificity, we define the radial secular trap frequency as $\omega_r\equiv \text{Max}[\omega_x,\omega_y]$ throughout. 
Doppler cooling of the ions is accomplished by irradiating the 369.5 nm $^2$S$_{1/2}\ket{F=0}\rightarrow ^2$P$_{1/2}\ket{F=1}$ and $^2$S$_{1/2}\ket{F=1} \rightarrow ^2$P$_{1/2}\ket{F=0}$ transitions; ions are imaged by capturing the fluorescence from these transitions on an EMCCD camera.

\begin{figure}[t!]
\includegraphics[width=\columnwidth]{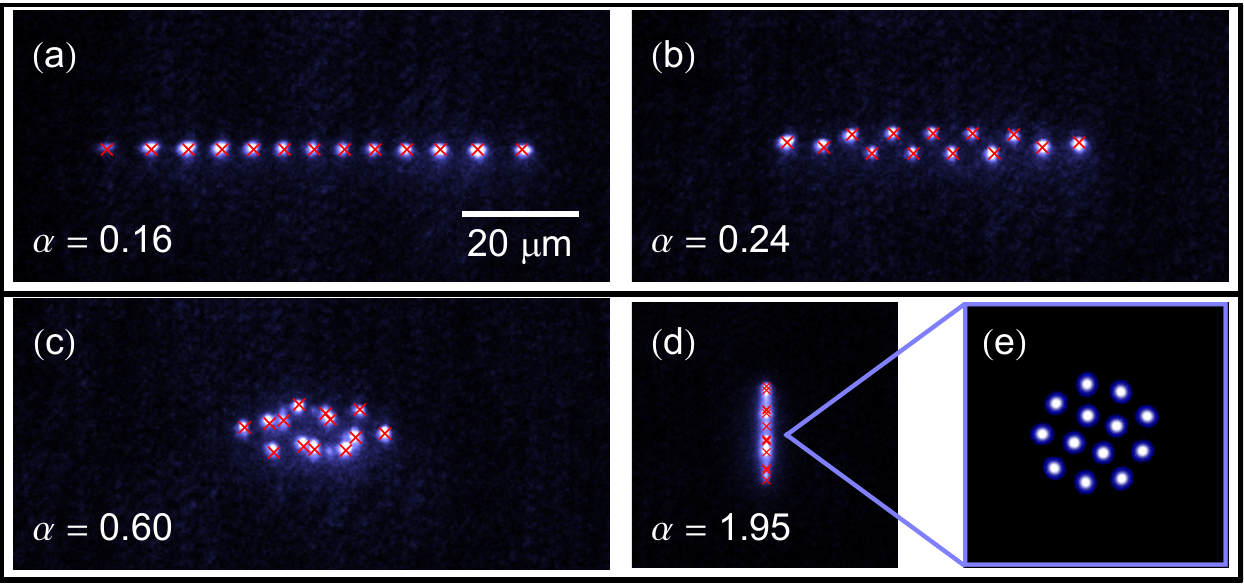}
\caption{Crystals of 13 ions are shown for increasing values of the trap aspect ratio $\alpha\equiv\omega_z/\omega_r$. The structure transforms from a 1D chain (a) to zig-zag (b) and 3D spheroidal phases (c), before ending in a 2D triangular lattice in the radial plane (d). Crosses show the ion positions predicted by the pseudopotential approximation. Panel (e) shows the same calculation as the crosses in (d), rotated to better display the lattice structure. Simulated ion sizes in (e) correspond to the diffraction-limited spot size of our imaging optics and include effects from rf-driven micromotion.}
\label{fig:Configurations}

\end{figure}

\emph{Structural Phase Transitions}---When the aspect ratio $\alpha\equiv\omega_z/\omega_r$ of the trap's axial and radial secular frequencies is small, ions form a 1D chain along the trap's central axis (Fig \ref{fig:Configurations}(a)). 
As $\alpha$ is increased (by increasing the axial frequency), the ions pass through a zig-zag phase (Fig. \ref{fig:Configurations}(b)) and a number of three-dimensional (3D) spheroidal configurations (Fig. \ref{fig:Configurations}(c)), before forming a radial-2D crystal. This last configuration occurs in Fig. \ref{fig:Configurations}(d), where the single plane of ions is viewed on-edge. Fig. \ref{fig:Configurations}(e) simulates the same crystal rotated perpendicularly to the plane. For these higher-$\alpha$ phases, ions that lie away from the trap's central axis are subject to rf-driven micromotion, the amplitude of which increases linearly with an ion's radial coordinate \cite{sm}. Though the equilibrium ion positions are no longer stationary due to micromotion, the observed time-averaged positions closely correspond to predictions obtained from pseudopotential theory calculations (red crosses in Fig. \ref{fig:Configurations}).

\begin{figure}[t!]
\includegraphics[width=\columnwidth]{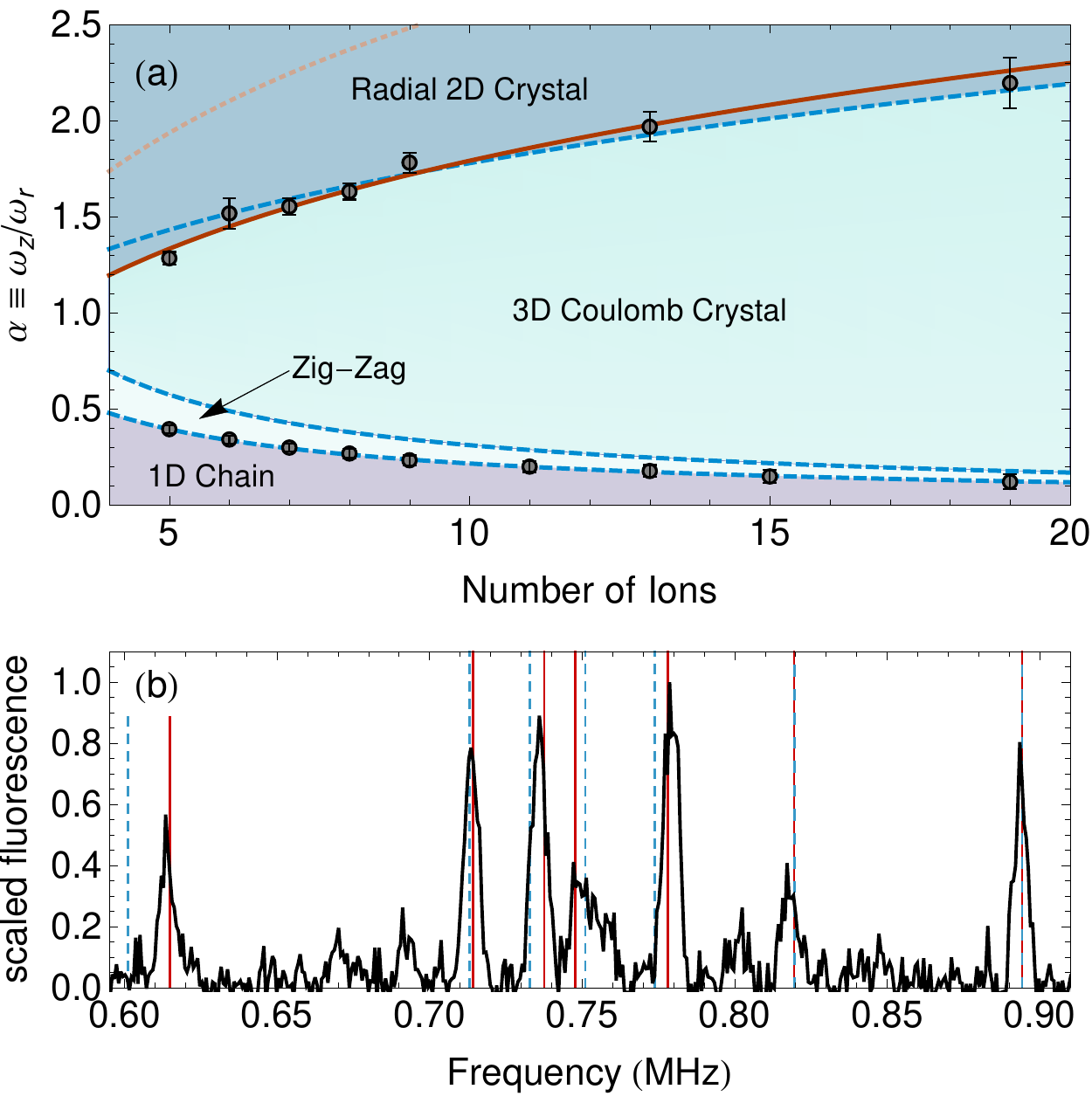}
\caption{(a) Phase diagram of ion Coulomb crystals in a linear Paul trap. Data show the measured $\alpha$ that separate the 1D/zig-zag and 3D/radial-2D phases as a function of ion number. Three theory predictions (with no adjustable parameters) are plotted for comparison. Blue dashed, pseudopotential \cite{dubin1993theory}; Red solid, Floquet-Lyapunov \cite{landa2012modes}; Orange dotted, micromotion-destabilized \cite{wang2015quantum}. (b) Axial mode spectrum for $7$ ions in a radial-2D crystal at $\alpha = 2$. Vertical lines show predicted mode frequencies. Blue dashed, pseudopotential; Red solid, Floquet-Lyapunov.}
\label{fig:Phase}

\end{figure}

Varying the axial confinement over such a large range enables the precise experimental determination of structural phase transition boundaries at both small and large $\alpha$, as shown in Figure \ref{fig:Phase}(a). 
Ions starting in a 1D chain exhibit a sudden transition to a zig-zag configuration at a critical value of $\alpha$ dependent on particle number $N$
\footnote{
The 1D to zig-zag transition only occurs due to non-degenerate radial frequencies and is quite close to the zig-zag to 3D boundary for our near-degenerate trap. Numerous unmapped subtransitions occur within the 3D Coulomb crystal phase; the richness of 3D geometries that arises with even 3-4 ions is detailed in \cite{kaufmann2012precise}
}.
Since micromotion plays no role in this transition, numerical estimates of the phase boundary are straightforward \cite{dubin1993theory,schiffer1993phase, steane1997ion} and have been previously verified with up to 10 ions \cite{enzer2000observation}.

Our measurements confirm this behavior for up to 19 ions and are compared to the theoretical prediction (lowest blue dashed line) in Fig \ref{fig:Phase}(a).

For the 3D to radial-2D transition, the presence of micromotion complicates theoretical estimates of the phase boundary. One calculation, shown as the upper blue dashed line in Fig. \ref{fig:Phase}(a), predicts the phase transition using only the time-averaged pseudopotential \cite{dubin1993theory}. A more complete description, which accounts for the fully-coupled and time-dependent dynamics of the ion positions, is shown as the solid red line in Fig. \ref{fig:Phase}(a). Here, a Floquet-Lyapunov (FL) transformation is invoked to convert the periodic, time-dependent problem to a time-independent formulation and find the decoupled modes of oscillation \cite{landa2012classical, landa2012modes}. 

A third analysis of this phase boundary, shown as the orange dotted line in Fig. \ref{fig:Phase}(a), suggests the existence of a micromotion-destabilized region due to a downward shift in transverse mode frequencies \cite{wang2015quantum}. Our measurements of the 3D to radial-2D phase boundary in Fig. \ref{fig:Phase}(a) confirm the validity of the FL approach in this regime, as opposed to the micromotion-destabilized theory. In addition, our data demonstrate that the pseudopotential approach provides a close approximation of the transition for up to 19 ions, even in the presence of increasing radial micromotion with larger crystal sizes.

As a further investigation of micromotion effects, we measure the vibrational spectrum of a 7-ion crystal deep in the radial-2D regime. Global, far-detuned Raman transitions at 355 nm allow for spin-motion coupling and coherent excitation of the crystal modes \cite{campbell2010ultrafast}. The two Raman beams have a frequency difference near the $^{171}$Yb$^+$ hyperfine ground state splitting $\omega_{\text{hf}}$, with the precise frequencies, amplitudes, and relative phases controlled by acousto-optic modulators \cite{smith2016many}. In our experiment, the wavevector difference of our Raman beams is aligned perpendicularly to the crystal plane, resulting in strong coupling to the axial (transverse) modes and suppression of coupling to the radial (in-plane) modes.

In Fig. \ref{fig:Phase}(b), we compare the measured axial mode frequency spectrum to frequencies calculated using the pseudopoential (blue dashed) and FL (red solid) approaches.
These methods largely agree with the measured data and with each other to within 2 kHz, though the pseudopotential approximation mispredicts the lowest frequency mode by over 10 kHz. Nevertheless, the pseudopotential approximation may still provide reasonable accuracy for many experiments. For instance, in quantum simulations of spin-lattice Hamiltonians \cite{monroe2021programmable}, the pseudopotential approach correctly predicts the 2D-Ising interaction range to within $0.5\%$ for up to 19 ions.


\emph{Rf heating effects---}The presence of micromotion may have strong effects on crystal lifetimes and temperatures. When multiple ions are confined in an rf trap, ion-ion collisions can transfer micromotion energy into secular kinetic energy and result in rapid rf heating \cite{ryjkov2005simulations,zhang2007molecular}.
As the collision rate increases, ion motion becomes less correlated, and a sudden jump in temperature occurs at an inflection point which corresponds to a `melting' of the crystal \cite{chen2013measurement}.
This rf heating mechanism is expected to dominate over other sources of noise, such as electric field fluctuations \cite{brownnutt2015ion} and collisions with background gas molecules \cite{buluta2008investigation}. Though molecular dynamics simulations indicate that large numbers of ions could be maintained for long times without continuous cooling \cite{buluta2008investigation}, this presumes the existence of ideal traps; no prior studies have established the lifetime and heating rates of radial-2D crystals in experimentally-realizable systems.

To begin investigating the effects of micromotion-induced heating, we measure the trapping lifetimes of radial-2D crystals in the absence of active cooling. 
After the ions are Doppler cooled, the cooling beams are switched off and the ions are allowed to heat for a specified amount of time.
If the crystal melts during this period, one or more ions may escape the trap confining potential or remain uncooled when the Doppler beams are re-applied.
We define the trapping lifetime as the time for which all ions remain in the crystal with 1/e probability, and find that it is in excess of 5 seconds for lattices of up to 19 ions. 
This lifetime is exceptionally long compared to the typical $\sim$millisecond timescales of quantum computation or simulation experiments \cite{richerme2014non,linke2017experimental}.

\begin{figure}[t!]
\includegraphics[width=\columnwidth]{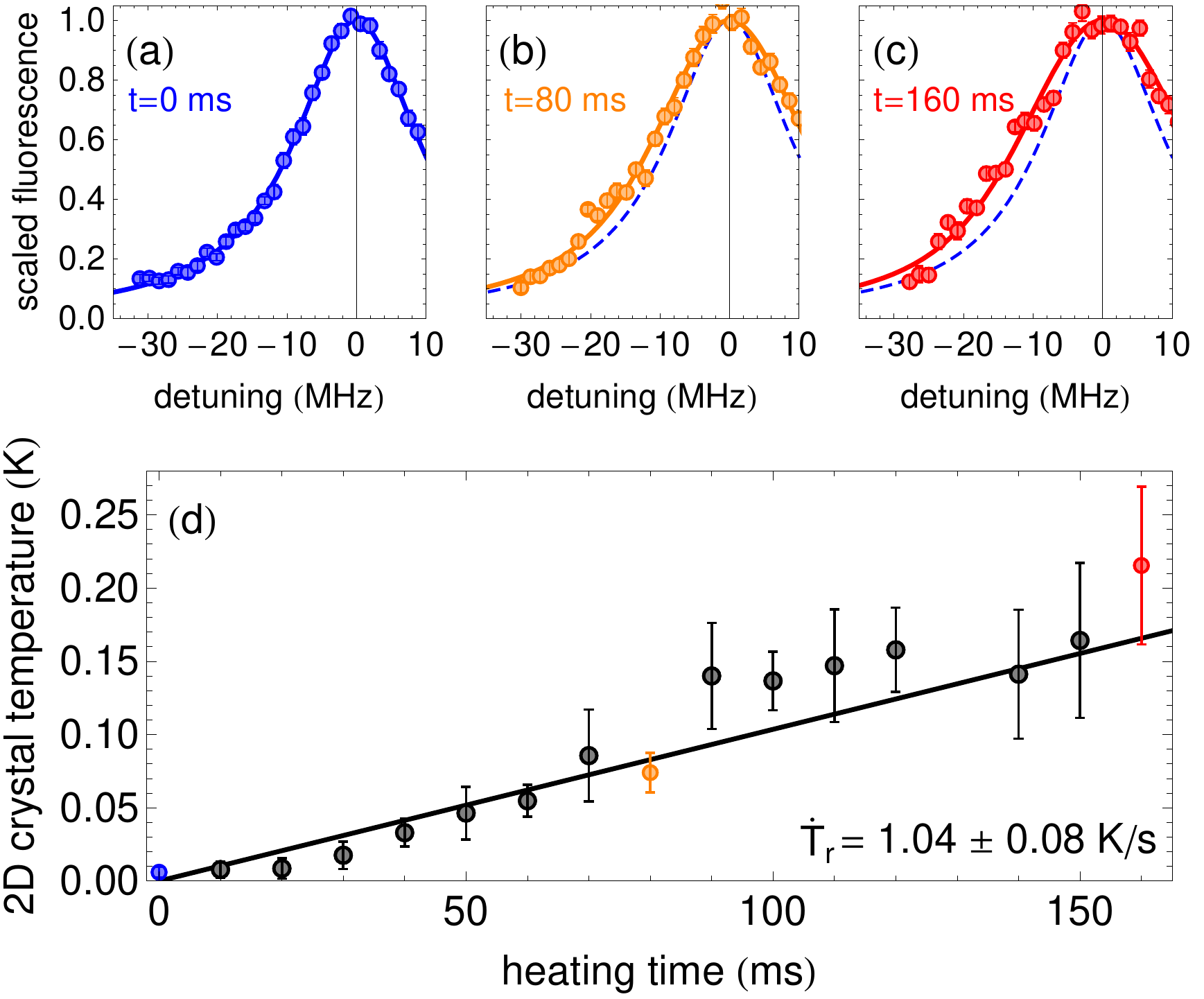}
\caption{Voigt fluorescence lineshapes of a 7 ion crystal exposed to rf heating are shown for heating times of (a) $0$ ms, (b) $80$ ms, and (c) $160$ ms. The lineshape widens at later times due to increasing contributions from Doppler broadening. The $0$ ms profile (3 mK temperature) is indicated by a dashed blue line in panels (b) and (c) for reference.
(d) A radial heating rate of $\dot{T}_r=1.04 \pm .08$ K/s is extracted from Voigt profile fits to ion fluorescence data.}
\label{fig:Voigt}
\end{figure}

To further study rf heating effects, we determine the temperature of the radial-2D crystal by analyzing the ions' fluorescence lineshape.
The ion resonance, which is described by a Voigt distribution, is a convolution of Lorentzian and Gaussian profiles. The Lorentzian contribution comes from the power-broadened natural linewidth $\Delta\nu_L=\Gamma\sqrt{1+s} = 2\pi\times 22$ MHz, where $\Gamma = 2\pi\times 19.6$ MHz is the natural linewidth of the $^{171}$Yb$^+$ 369.5 nm $^2$S$_{1/2} \rightarrow ^2$P$_{1/2}$ transition and $s=0.3$ is the laser saturation parameter. The Gaussian contribution results from Doppler broadening, with a full-width at half-maximum of $\Delta\nu_G = 2\sqrt{\frac{(2\ln 2) k_B}{m\lambda^2}}\sqrt{T_r\cos^2\theta+ T_z\sin^2\theta}$. This expression arises since our fluorescence beam intersects the crystal plane at an angle ($\theta=45^{\circ}$) and is therefore sensitive to both the radial and axial temperatures $T_r$ and $T_z$. Later we will show that keeping independent radial and axial temperatures is well-justified, and that the axial temperature adds negligible contribution to the overall linewidth.

To extract the radial crystal temperature, we fit the measured Voigt fluorescence profile to a Lorentzian of constant width $\Delta\nu_L$ and a Gaussian of variable width $\Delta\nu_G$. When the crystal is Doppler cooled to 3 mK (as confirmed with sideband Raman spectroscopy), the Gaussian contribution is small and the line profile is essentially Lorentzian (Fig. \ref{fig:Voigt}(a)). However, if the cooling beams are extinguished and the crystal acquires radial energy through rf heating, the fluorescence profile spreads due to an increase in thermal motion (Fig. \ref{fig:Voigt}(b,c)). By performing many temperature measurements at increasing heating times, as shown in Fig. \ref{fig:Voigt}(d), we determined the radial heating rate to be \mbox{$\dot{T}_r=1.04 \pm .08$ K/s}. Previous work has predicted nonlinear heating near the melting point of Coulomb crystals; the linear nature of our data implies that short time scales, large ion masses, and low initial temperatures keep crystals far from this limit \cite{chen2013measurement, zhang2007molecular}.

\begin{figure}[t!]
\includegraphics[width=\columnwidth]{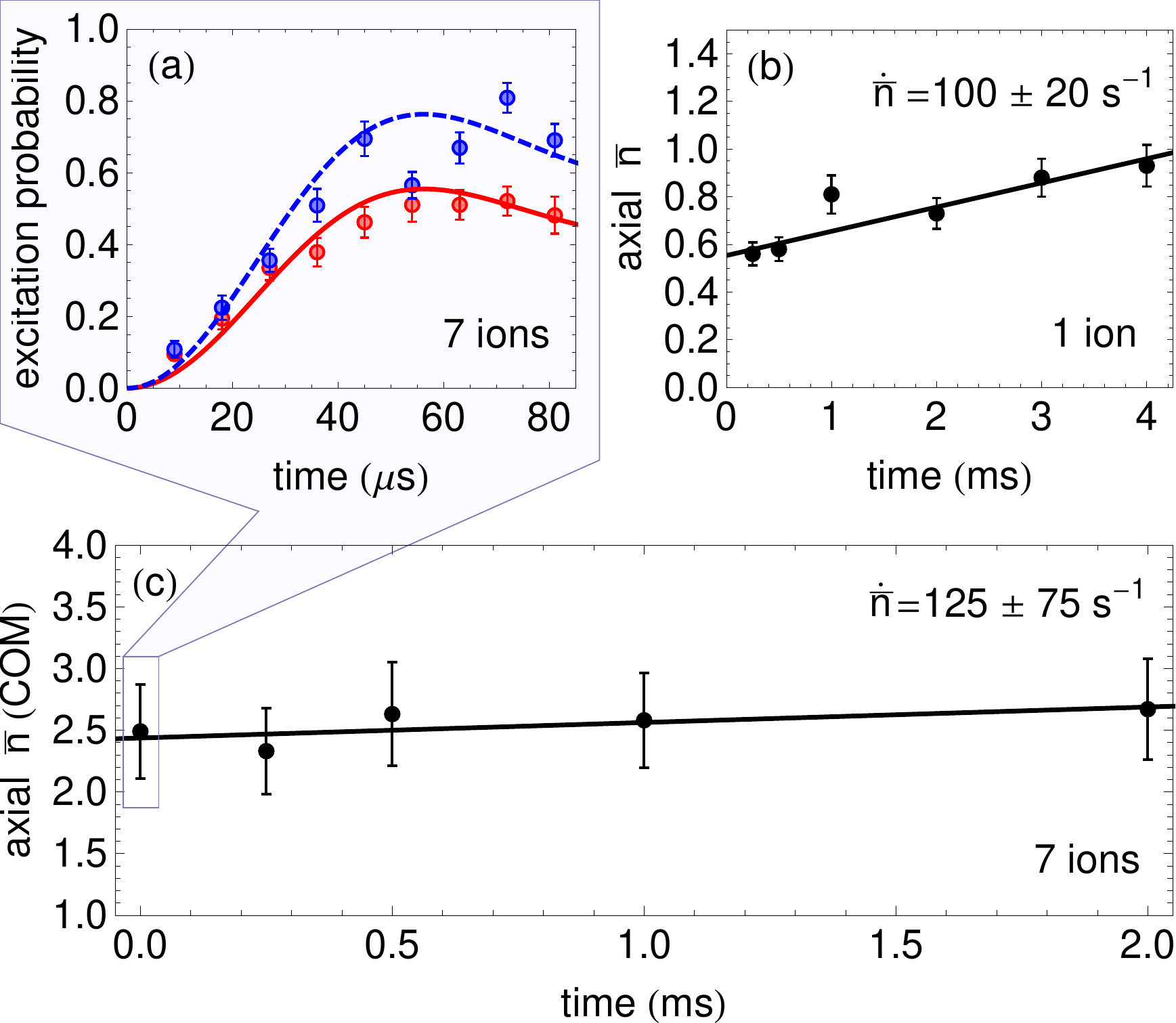}
\caption{(a) A comparison of red (solid) and blue (dashed) sideband probability amplitudes are shown for a 7 ion crystal immediately following sideband cooling. The heating rate of the axial (transverse) COM mode for a single ion (b) is comparable to that of a 7 ion crystal (c). In both cases, the absolute heating rate is low compared to traps of similar size.}
\label{fig:AxialHeating}
\end{figure}

To look for evidence of heat transfer between the radial and axial directions, we measure the heating rate of the axial center-of-mass (COM) mode using resolved sideband spectroscopy \cite{wineland1998experimental}. Following Doppler cooling, our 355 nm Raman beams are used to sideband cool the axial COM mode to $\bar{n} \lesssim 2.5$ as well as to induce stimulated Raman transitions at the axial COM red and blue sideband frequencies, $\omega_{\text{hf}} \pm \omega_z$. 

The number of quanta in the axial COM mode is determined by taking the ratio $r$ of red to blue sideband transition probability amplitudes (Fig. \ref{fig:AxialHeating}(a)) for several different sideband drive times and finding the mean occupation number $\bar{n} = \frac{r}{1-r}$. 

Finally, the axial COM heating rate $\dot{\bar{n}}$ is determined by leaving the crystal uncooled for increasing time periods and repeating the sideband measurements.

We compare the axial COM heating rate of a single ion to that of a radial-2D crystal with 7 ions, under the same trapping conditions (\mbox{$\omega_z \approx 2\pi\times 900$ kHz} and $\alpha=2$).

As shown in Fig. \ref{fig:AxialHeating}(b), we find a single-ion ambient heating rate of $\dot{\bar{n}}=100 \pm 20$ motional quanta/s. This measurement, which corresponds to temperature heating rate $\dot{T}_z=0.004\pm0.001~$K/s and a spectral density of electric field noise \mbox{$S_E=2.65\times10^{-12}$ V$^2$m$^{-2}$Hz$^{-1}$}, is comparable to heating rates observed in other room-temperature rf traps of similar size \cite{brownnutt2015ion}.
We then repeat these measurements for the axial COM mode of a 7-ion crystal, finding a heating rate of $\dot{\bar{n}}=125 \pm 75$ quanta/sec (Fig. \ref{fig:AxialHeating}(c)).

In temperature units, this rate is over 200 times smaller than the measured radial heating (Fig. \ref{fig:Voigt}) and justifies our earlier assumption of non-equilibration between axial and radial directions. 

Our measurements with one and seven ions further suggest that electric field noise is not the dominant heating mechanism in our trap. This is because electric field fluctuations, which are largely correlated across the ions, are expected to preferentially heat the COM mode and scale linearly with ion number \cite{brownnutt2015ion}. Our results instead indicate largely uncorrelated noise, which has likewise been observed in Penning traps using the analog of a radial-2D crystal \cite{sawyer2014spin}. In the limit of perfectly uncorrelated noise, we would expect other axial modes (indexed by $k$) to exhibit heating rates $\dot{\bar{n}}_{(k)}=(\omega_\text{COM}/\omega_k)\dot{\bar{n}}_{\text{COM}}$ \cite{brownnutt2015ion}, giving at worst an estimated $\sim 50\%$ larger heating rate for the lowest-frequency (zig-zag) axial mode. Whether the noise in our system is correlated or not, our observations of objectively low axial temperatures in the presence of rapid radial heating demonstrate that the axial modes of a radial-2D crystal remain cold, isolated, and well-suited for quantum simulation experiments.

\emph{Discussion and Outlook}---
Our experiments establish that micromotion effects on radial-2D crystals are largely constrained to the radial plane: phase boundaries and axial vibrational spectra are well-predicted by micromotion-free pseudopotential theory, and only the in-plane radial degrees of freedom experience micromotion-induced heating. In contrast, the axial (transverse) degrees of freedom remain decoupled and cold. Furthermore, we have enacted $\sim$5-$\mu$m ion-ion spacings in this geometry, which will enable fast ion-ion coupling rates while allowing for future individual addressing with low cross-talk.

Our demonstration of stable, isolated, and low-noise axial modes establishes radial-2D crystals in linear Paul traps as a realistic platform for implementing several proposals in quantum simulation \cite{richerme2016two, yoshimura2015creation}.
This system is especially well-suited for studies of highly-frustrated quantum spin models \cite{balents2010spin,richerme2016two,nath2015hexagonal,diep2013frustrated}, since long-range antiferromagnetic interactions are routinely implemented between co-trapped ions \cite{monroe2021programmable}, and since ions in the radial-2D phase self-assemble into a triangular lattice. Using only global laser beams, we will be able to characterize the ground state and dynamical properties of frustrated 2D spin-models by measuring their excitations \cite{balents2010spin} and correlation functions (which can distinguish, for instance, between N\'eel states or Valence Bond Solid states \cite{lhuillier2005frustrated}), and by tuning the relative contributions of inherent geometric and long-range frustration.

Realization of such proposals with radial-2D crystals will demand several future developments. First, the imaging optics should be moved perpendicularly to the crystal plane to facilitate site-resolved detection of the ion lattice. Next, methods to cool radial-2D crystals near the motional ground-state should be applied, as they have been for lateral-2D crystals \cite{qiao2021double}. Evidence of entanglement generation via M\o lmer-S\o rensen interactions \cite{molmer1999multiparticle} (or equivalent) should then be demonstrated before implementing full spin-lattice simulations. Finally, the possibility of maintaining 100+ ions in the radial-2D crystal phase for long times \cite{buluta2008investigation}, and the limits of crystal stability in the presence of rf heating, should be experimentally explored as the system is scaled to larger sizes.

The possibility to perform individual ion addressing in rf traps, which is already well-established for 1D ion chains \cite{smith2016many,linke2017experimental,wright2019benchmarking}, will further expand the capabilities of the radial-2D platform. Shelving of specific ions will allow for the quantum simulation of more complex lattice geometries, such as Kagome, which are believed to support spin-liquid phases \cite{richerme2016two,balents2010spin,nath2015hexagonal,yan2011spin,kumar2015chiral}. 

Furthermore, radial crystals with individual addressing could provide a naturally scalable solution for fault-tolerant quantum computing \cite{wang2015quantum,shen2014high} or simplify preparations for one-way quantum computing schemes \cite{raussendorf2001one, wunderlich2009two}.

\begin{acknowledgments}
This work was supported by the U.S. Department of Energy, Office of Science, Basic Energy Sciences, under Award $\#$DE-SC0020343. The IU Quantum Science and Engineering Center is supported by the Office of the IU Bloomington Vice Provost for Research through its Emerging Areas of Research program.
\end{acknowledgments}

\bibliographystyle{prsty}
\bibliography{main}{}

\onecolumngrid
\newpage
\begin{center}
\textbf{\large Supplementary Material for \\
``Radial two-dimensional ion crystals in a linear Paul trap" \\~}
\end{center}
\twocolumngrid




\title{Supplementary Material for \\
``Radial two-dimensional ion crystals in a linear Paul trap"}
\date{\today}

\maketitle

\section{Trapping Potential and Secular Frequencies}

Near the center of a linear Paul trap, the time-dependent potential is written \cite{wineland1998experimental}
\begin{equation}
    V(\vec{r}, t) = \frac{V_0 \cos(\Omega_{t} t)}{2 r_0 ^2} (x^2-y^2) + \frac{\kappa U_0}{2 z_0^2}(2z^2-x^2-y^2)
\end{equation}
where $V_0$ and $U_0$ are the rf and dc voltages, $r_0$ and $z_0$ are the radial and axial trap dimensions, $\Omega_{t}$ is the trap drive frequency, and $\kappa$ is a geometric factor of order 1. 
Equations of motion in the $x$ and $y$ directions for an ion confined in this potential are described by the standard Mathieu equations 
\begin{equation}
    \frac{d^2 u_i}{d \zeta ^2} + [a_i + 2 q_i \cos(2 \zeta)] u_i = 0
\end{equation}
for directions $i \in \{x, y\}$, dimensionless time $\zeta = \frac{\Omega_t t}{2}$, and Mathieu parameters 
\begin{equation}
a_{x,y} = -\frac{4Q \kappa U_0}{m z_0^2 \Omega_t^2} \text{ , } q_x = -q_y = \frac{2QV_0}{m r_0^2 \Omega_t^2}. 
\end{equation}
for ion mass $m$ and charge $Q$.

For $a_i < q_i^2 \ll 1$, the solution (to first order in $a$ and second order in $q$) is given by 
\begin{equation}
\label{ui}
\begin{split}
    u_i(t) = A_i \bigg( \cos(\omega_i t) \left[1+\frac{q_i}{2}\cos(\Omega_t t)+\frac{q_i^2}{32}\cos(2 \Omega_t t) \right] \\
    +\beta_i \frac{q_i}{2}\sin(\omega_i t) \sin(\Omega_t t)\bigg)
\end{split}
\end{equation}
for $\beta_i \approx (a_i + q_i^2/2)^{1/2}$ and $A_i$ dependent on initial conditions.
The large amplitude motion at $\omega_i$ is the secular motion, with resonant frequencies $\omega_i = \beta_i \Omega_t /2$, while the fast, small oscillation at $\Omega_t$ is known as micromotion.

For low temperatures and weak axial confinement, the secular motion of an ion in a radially-symmetric trap is well-described by the 3D harmonic pseudopotential
\begin{equation}
    Q \Phi(r, z) = \frac{1}{2} m \left(\omega_r ^2 r ^2 + \omega_z ^2 z ^2\right)
\end{equation}
where $r^2=x^2+y^2$. In this approximation, the trap secular frequencies can be written
\begin{equation}
    \omega_{r} = \sqrt{\frac{Q}{m}\left(\frac{qV_0}{4 r_0^2}-\frac{\kappa U_0}{z_0^2}\right)} \text{ , } \omega_{z} =  \sqrt{\frac{Q}{m}\frac{2\kappa U_0}{z_0^2}}.
\end{equation}
In our experiments with radial-2D crystals, a slight asymmetry is introduced between the $x$ and $y$ directions to break the radial degeneracy and prevent a zero-frequency rotational mode. In the main text, we choose  $\omega_r\equiv \text{Max}[\omega_x,\omega_y]$ for specificity.\\

\begin{figure}[t!]
\includegraphics[width=\columnwidth]{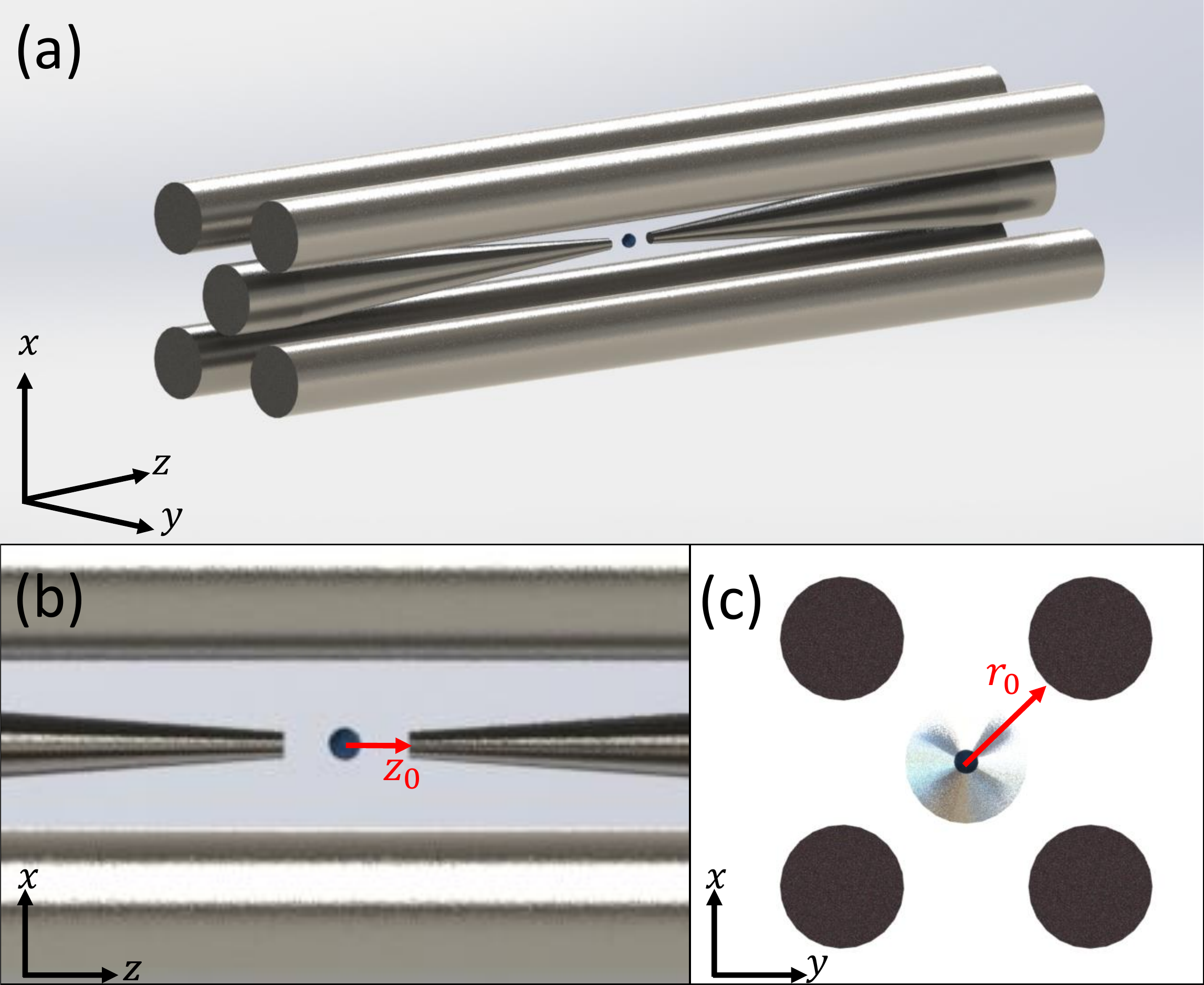}
\caption{(a) Full view, (b) front view, and (c) side view of the four-rod linear Paul trap used for the experiments described in the main text, with designated trap coordinate axes (lower left) and dimensional parameters (red).}
\label{fig:rodtrap}
\end{figure}

\section{Trap Geometry}

Experiments are performed with $^{171}$Yb$^+$ ions confined in a four-rod linear Paul trap (Fig. 1) with two ``needle" endcaps along the $\hat{z}$ (axial) direction.
An oscillating rf voltage $V_0$ is applied to two opposing rods (the other two are grounded), creating a quadrapole potential in the $xy$ (radial) plane, while a static dc voltage $U_0$ is applied to the two endcaps.
Radial and axial trap dimensions are given by $r_0=460~\mu$m and $z_0=335~\mu$m, respectively, and the trap drive frequency is \mbox{$\Omega_{t}=2\pi\times21$ MHz}.
The rf voltage $V_0$ is held constant at $340$ V throughout experiments, yielding a Mathieu parameter of $q = .10$, while the dc voltage $U_0$ is varied between $.012-30$ V. The trap geometric factor is determined through frequency measurements to be $\kappa \simeq 0.12$.
As the dc voltage is raised, the aspect ratio $\alpha = \omega_z/\omega_r$ increases. At large values of $\alpha$, ions are squeezed into the $xy$ plane and self-assemble into a triangular lattice.

Two $355$ nm Raman beams oriented in the $yz$ plane enter from the front of the trap, at $53$ degree angles to the left and right of the $y$ axis respectively. For the sideband cooling preparation and heating rate measurements in Fig. 4 of the main text, for which $\omega_z = 2\pi\times 900$ kHz, this leads to a dimensionless Lamb-Dicke parameter 
\begin{equation}
    \eta = 2 \sin(53^\circ)\frac{2\pi}{\lambda}\sqrt{\frac{\hbar}{2m\omega_z}} = 0.16
\end{equation}
and a resultant wavevector $\vec{k} = \vec{k_2} - \vec{k_1}$ along the axial ($\hat{z}$) direction. For a radial-2D crystal, oriented in the $xy$ plane, this geometry results in strong coupling to the axial (transverse) motional modes and suppression of coupling to the radial modes.

\section{Micromotion Amplitude}

The time-independent potential experienced by a radial-2D crystal of $N$ ions in the $xy$ plane is the combination of the trapping potential and the Coulomb potential, written

\begin{multline}
   V(x, y) = \sum_i (\frac{1}{2} m \omega_x^2 x_i^2 + \frac{1}{2} m \omega_y^2 y_i^2) \\
   + \sum_{i < j} \frac{Q^2}{4 \pi \epsilon_0 \sqrt{(x_i-x_j)^2 + (y_i-y_j)^2}} 
\end{multline}
for $i = 1, 2, ... N$.

The equilibrium ion positions $\vec{r}_0 = (x_{0}, y_{0})$ are the set of coordinates where the system's energy is minimized; each ion lies a distance $r_0$ away from the trap's central axis. Following Eq. \ref{ui} above, micromotion drives the ions' radial coordinates about their equilibrium positions as 
\begin{equation}
    \vec{r}(t) = \vec{r}_0 + \vec{r}_1 \cos(\Omega_t) + \vec{r}_2 \cos(2 \Omega_t) + ...
\end{equation}
with coefficients $|\vec{r}_1| = \frac{q}{2}r_0$ and $|\vec{r}_2| = \frac{q^2}{32}r_0.$ 
Since $q = 0.10$ for our trap parameters, the micromotion amplitude is sufficiently well approximated by $|\vec{r}_1|$.

In the main text, axial mode frequency and heating rate measurements are performed with a seven ion crystal, where the outermost ions are approximately 6 $\mu$m from the trap center. For this crystal geometry, we find that the micromotion amplitude for these ions is \mbox{$\approx 300$ nm}. For the $13$-ion crystal shown in Fig. 1, we find $|\vec{r_1}| \approx 430$ nm, and for the largest examined crystal of $19$ ions, we find $|\vec{r_1}| \approx 500$ nm. In each case, we note that the micromotion amplitude is  small compared to the ~$5~\mu$m inter-ion spacing of the crystal. As the number of ions $N$ grows larger, the outer radius of the crystal scales as $\sim d\sqrt{N}/2$ for ions separated by distance $d$. The maximum micromotion amplitude therefore also scales with the square-root of ion number, $|\vec{r_1}|_\text{max} \sim qd\sqrt{N}/4$ \cite{richerme2016two}.
\onecolumngrid

\widetext

\end{document}